\documentclass[sigconf]{acmart}
\renewcommand\footnotetextcopyrightpermission[1]{} 
\usepackage{booktabs} 
\usepackage{flushend}
\usepackage[ruled]{algorithm2e} 

\usepackage{subfigure}
\usepackage{enumitem}

\setcopyright{none}

\acmArticle{4}
\acmPrice{15.00}

\begin{document}

\title{PUTWorkbench: Analysing Privacy in AI-intensive Systems}

\author{Saurabh Srivastava}
\affiliation{%
  \institution{Indian Institute of Technology, Kanpur}}
\email{ssri@iitk.ac.in}
\author{Vinay P. Namboodiri}
\affiliation{%
 \institution{Indian Institute of Technology, Kanpur}}
\email{vinaypn@iitk.ac.in}
\author{T.V. Prabhakar}
\affiliation{%
  \institution{Indian Institute of Technology, Kanpur}
}
\email{tvp@iitk.ac.in}

\begin{abstract}
AI intensive systems that operate upon user data face the challenge of balancing data utility with privacy concerns. We propose the idea and present the prototype of an open-source tool called Privacy Utility Trade-off (PUT) Workbench which seeks to aid software practitioners to take such crucial decisions. We pick a simple privacy model that doesn't require any background knowledge in Data Science and show how even that can achieve significant results over standard and real-life datasets. The tool and the source code is made freely available for extensions and usage.
\end{abstract}

%
%
  
%
%



\maketitle


\section{Introduction}
The recent advancements in Machine Learning techniques has brought AI experts and Software Practitioners on the same page. There is however a dearth of tools specially designed for building AI intensive systems. Such tools would be conceptually different from the conventional IDEs since they'll primarily focus on \textit{models}, instead of \textit{programs} or \textit{modules}. Also, a greater amount of time would be spent on \textit{fine-tuning} existing models and finding the right \textit{parameters} to a learning algorithm, instead of building exhaustive test suites to test handwritten code. Another important difference would be that instead of using a binary logic in evaluation of the models (e.g. a test case that either passes or fails), we'll need to have mechanisms to deal with probabilities (e.g. a Classification Accuracy of 89.2\%). This provides enough motivation to come up with novel ideas about newer tools, dealing with problems that arise in systems dealing with these paradigms.
\subsection{Privacy-Utility Trade-off}
The starting point of any Machine Learning activity includes collecting or acquiring a significant quantity of data, the more, the merrier. More often than not, this data belongs to individuals who are protected by legal guidelines such as EU's GDPR\cite{regulation2016regulation}, putting additional tasks on software providers to ensure that they do not violate their privacy. The \textit{utility} of a learning algorithm depends on its ability to ``find correlations'' among the provided attributes. However, not all correlations may be desirable, as they may lead to revealing the identity of individuals who were assured that their data would be sanitised via anonymisation before use. An example of how this can be done from medical records, in conjunction with a voter list, is shown in \cite{samarati2001protecting}. In any case, anonymising the data, or any other similar activity performed to achieve higher levels of \textit{privacy}, usually involve ``breaking correlations'' that can lead to user identification. The exact correlations that need to be preserved or broken depend on the actual use-case. In essence, any Machine Learning activity, when applied in a real-world software system, needs to achieve a \textit{trade-off} between privacy and utility. A tool that allows a software practitioner to evaluate custom privacy scenarios against their datasets can help achieve a near optimal solution on this virtual trade-off scale.
\subsection{Envisioned Users}
The users who we envision to pay attention to such a tool are software practitioners who are working on a new AI-intensive system or those who are tasked with ensuring that a similar existing system meet certain privacy guidelines. For instance, the GDPR\cite{regulation2016regulation} compliance date of May 25, 2018 forced many providers to send revised privacy notices to their users as they'll be liable for legal action if they do not meet the regulation's requirements beyond the deadline. We can expect that in future, more similar guidelines will be put in place, with which the software providers will need to abide. One particular challenge that hampers adoption of sophisticated privacy techniques is that more often than not, it'll require an expert in the field of Data Science to understand the nuances of complex correlations and befuddling statistical metrices. A typical software practitioner need not necessarily have the expected skill-set to deal with this scenario. Our primary goal is to build a tool that doesn't require upfront knowledge of these concepts. The initial version uses a model that doesn't involve any extra metrices than what the practitioner may have already seen.
\begin{figure*}[ht]
\centering
\includegraphics[width=\textwidth]{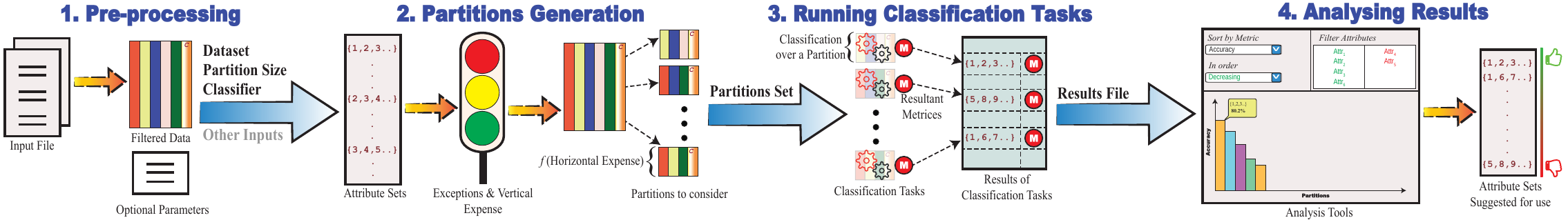}
\caption{The Methodology of PUTWorkbench Tool}
\label{fig:methodology}
\end{figure*}
\section{The Simple Privacy Model}
To build the initial version of the tool, we've used a model that is relatively simple and intuitive. The model is based on the \textit{Privacy as the Default Setting} principle\cite{cavoukian2009privacy}, specifically, relating to the FIPs \textit{Collection limitation} and \textit{Data minimization}\cite{cavoukian2009privacy}. In simple terms, if we don't keep something with us, we need not worry about it being stolen or exposed.

Take the example of a system that recommends products for a user on an e-commerce website. Assume that the website stores many attributes relating to its users. For instance, the time they last visited the site, the number of products they clicked on, the number of times they pressed the ``Back'' button, etc. This dataset is then used by a Recommender component to show users with relevant products. Let us say the \textit{utility} of the Recommender is measured in its prediction accuracy, i.e. the average percentage of instances where the user clicked on the recommended product.

Assume that the dataset contained \textit{n} attributes, and \textit{m} rows. Assume that the Recommender achieved an accuracy of $x\%$ with this dataset. Now consider a case when we randomly remove half of the attributes (and may be some of the rows too) from the dataset, i.e. we come up with a dataset that is a partition of the original dataset. Assume that we fed this dataset to the same Recommender, and it achieved an accuracy of $y\%$. Let us concentrate on the pessimistic case where this reduction in the size of the dataset hampered the utility of the Recommender, i.e. $y < x$.  What if the difference $(y - x)$ is not substantial? What if we are ``ok'' to lay off some percentage of utility, in exchange for a scenario where we store much less user information than we already do?

The Simple Privacy Model is based on this trade-off phenomenon. Based on a particular use-case, a team of Practitioners might actually choose a point where the achieved utility balances out the achieved privacy from the business perspective. Clearly, this model is simple enough to understand, without looking up a Statistics cheat-sheet. We choose this model to implement the first iteration of our tool since it is easy to understand.
\begin{figure*}[t]
\centering
\includegraphics[width=\linewidth]{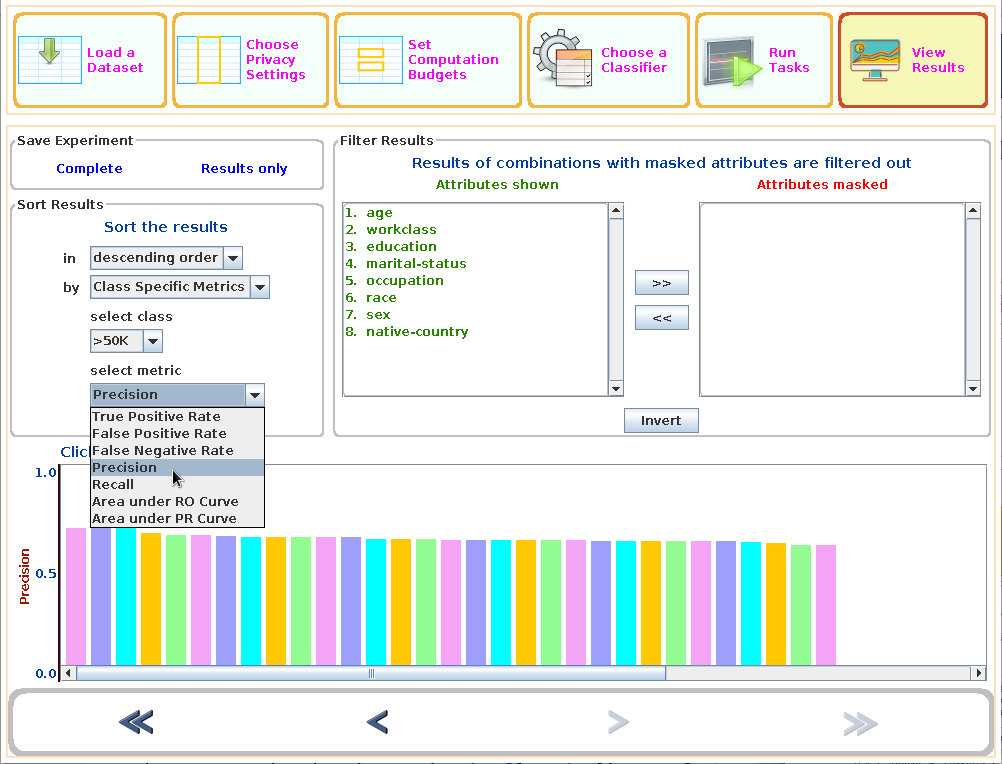}
\caption{The Analysis Tab of PUTWorkbench Tool}
\label{fig:tool}
\end{figure*}
\subsection{Core Parameters}
The model has the following core parameters:\\
\underline{Partition Size}:- The number of attributes to pick out of the original dataset. If there are $n$ attributes\footnote{Whenever we say attributes, we exclude the ``class'' attribute, unless explicitly specified } in the dataset, partition size is an integer between $1$ and $n$, both inclusive. Another way to provide this number is on a \textit{relative floating scale} between $-1.0$ and $1.0$, where $-1.0$ corresponds to a partition size of $1$, and $1.0$ corresponds to a partition size of $n$. On this scale, $0$ means $ceil(n/2)$. This scale is called \underline{PUTNumber} scale, where PUTNumber stands for ``Privacy Utility Trade-off Number'' which maps to a unique Partition Size.\\
\underline{Learning Objective}:- The purpose for which the dataset will be used. For the initial versions of the tool, we've included a small set of Classification techniques\cite{kotsiantis2007supervised} as Learning Objectives.\\
\underline{Privacy Exceptions}:- A known set of attribute sets which when put together, pose a higher risk to privacy (because they can induce undesirable correlations). Any partition that has a privacy exception, \textit{will not} be considered by the tool. \\
\underline{Utility Exceptions}:- A knows set of attribute sets which when put together, provide higher levels of utility (because they can induce desirable correlations). Any partition that has a utility exception, \textit{may be} considered at higher priority by the tool, in cases where it cannot consider all possible combinations.

The Partition Size and Learning Objectives are required parameters, while the exceptions are optional. It may be noted that for partitions which contain at least one privacy exception as well as one or more utility exceptions, privacy exceptions take precedence (i.e. such a partition is \textit{not considered} by the tool).  In essence, while privacy exceptions are ``bindings'', utility exceptions are mere ``suggestions'', which may or may not be accepted by the tool.

\subsection{Tool-Specific Parameters}
The possible number of partitions of size $k$, for a dataset with $n$ attributes is $^n$C$_k$. Even for moderately high values of $n$ and $k$, the number may easily be in millions. Also, number of rows in a dataset can range from $1$ to any arbitrarily large number. To make the tool computationally practical, we added some additional parameters:\\
\underline{Vertical Expense}:- A floating point number between $0$ and $1.0$, instructing what proportion of the overall possible partitions shall the tool consider. A Vertical Expense of $0.25$ means, ``try only 25\% of all possible combinations''.\\
\underline{Horizontal Expense}:- A floating point number between $0$ and $1.0$, instructing what proportion of the overall rows shall be part of any given partition. A Horizontal Expense of $0.10$ means, ``put only 10\% of total rows in any partition''.\\
\underline{Generation Method}:- Suggests a way to generate the Attribute-sets. \textit{Random Generation} method generates sets in a random fashion. \text{Dictionary Generation} method generates them systematically in a dictionary order\footnote{Dictionary order looks like this -> \{1,2,3\}, \{1,2,4\} ... \{1,2,n\}, \{2,3,4\}, \{2,3,5\} ...}. \\
There are some other lesser important parameters too in the tool which are meant for fine-tuning that we omit here. The default value for the two expenses is $1.0$. By default, the generation takes place in Dictionary Order.
\section{Methodology}
Figure \ref{fig:methodology} shows a pictorial overview of the tool. The overall methodology can be summarized as: \\
\begin{enumerate}[leftmargin=*]
    \item Runs a set of validity checks over the supplied parameters. For instance, it checks that the dataset has a ``nominal'' (and not ``numeric'') class attribute.
    \item Applies any filtering options over the dataset, if requested. For instance, removal of rows with missing values, or filling them with mean or mode.
    \item Decides upon the Attribute-set Generation Method. If the partitions to generate are fairly high, using the Random Method is generally a better option. The default method is Dictionary Method. The actual number of partitions to generate roughly equals ``$v$ times $^n$C$_k$'', where $v$ is the Vertical Expense, $n$ is the total number of attributes in the dataset, and $k$ is the Partition Size.
    \item Any Partitions containing a Privacy Exception are not considered by the tool any further. If $v$ was less than $1.0$, and some Partitions needed to be ignored, the one having Utility Exceptions are sparred the axe, to whatever extent possible.
    \item For each Attribute-set, a Partition of the dataset is created by including some or all rows from the original dataset, keeping only those attributes, which are in the said Attribute-set. The number of rows included in any partition roughly equals ``$h$ times $m$'', where $m$ is the total number of rows in the dataset. These rows are randomly selected from the original dataset.
    \item For each Partition, a Classification task is forked, using the requested technique. All the metrics produced by the task, such as Accuracy, False Positive Rate, Area under ROC Curve etc. are recorded.
    \item A Result file is prepared keeping all the above collected statistics. In the GUI version of the tool, a tab for analysing these results is shown, with multiple filtering options.
\end{enumerate}
At the core, the aim is to suggest the user with a subset of the attributes in the original dataset, which can provide significant utility for the selected Classification activity.
\begin{figure*}[ht]
\centering 
    \subfigure[J48 Classifier]{
        \label{fig:results-adult-iyengar-j48}
        \includegraphics[width=0.47\textwidth]{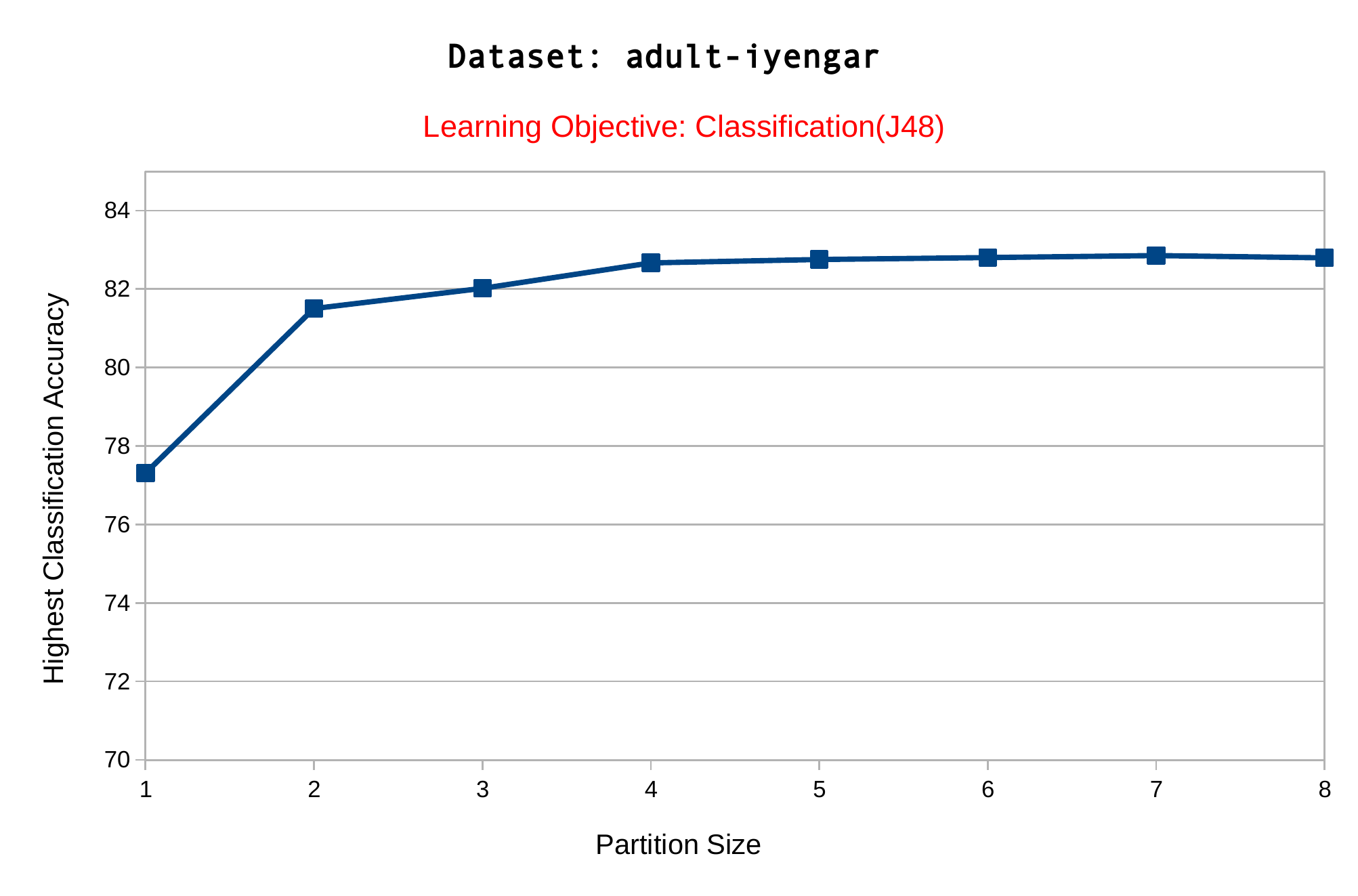}
    }
    \subfigure[Naive Bayes Classifier]{
        \label{fig:results-adult-iyengar-naivebayes}
        \includegraphics[width=0.47\textwidth]{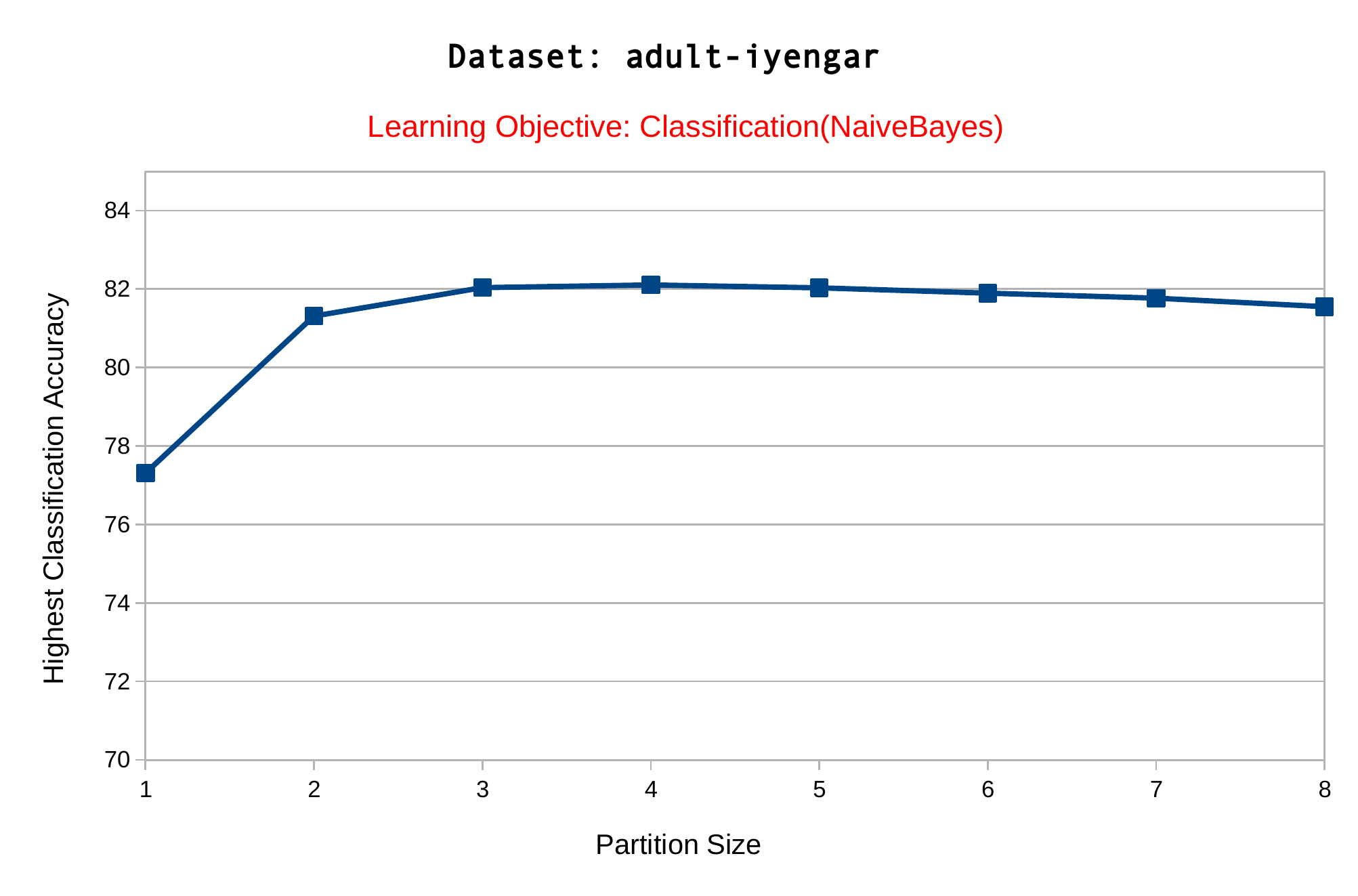}
    }
\caption{Highest Classification Accuracy achieved for \texttt{adult-iyengar} for varying values of partition size}
\label{fig:results-adult-iyengar}
\end{figure*}
\section{Tool}
The tool comes in two forms, one with a Command-Line Interface (CLI) and another with a GUI. The tool's package also contains two auxiliary helper tools. The tool is built using maven\cite{mavenurl} and is available as a public repository\cite{ssri2018} under the MIT license\cite{open2006license}. The \textit{Downloads} section of the repository provides zipped versions for Windows and Linux platforms, along with a packaged, JAR version. There is also a short User Manual accessible on the same page. An introductory video of the tool is available on YouTube\cite{ssri2018-vid}.

\subsection{GUI Version}
The UI version provides basic filtering and analysis facilities inline after the experiment completes. Figure \ref{fig:tool} gives a glimpse of the Analysis Tab of the UI tool. One can sort results according to a metric of choice, in increasing or decreasing order. Also, if one needs to see how well or poorly a set of attributes performed, that too can be achieved using the filtering options.
For the first time users, the GUI tool has an ``experimental'' feature called the \textit{Autopilot}. After loading a dataset, the Autopilot can be engaged to find common values of the required input parameters for the dataset. The Autopilot can be disengaged at any point in time if the user feels like taking control of the settings.
\subsection{CLI Version}
The CLI version can be handy if one needs to try out a spectrum of different partition sizes together. The CLI tool has switches to provide details of the dataset, Classifier, Partition Size, Exceptions, Expenses and several other settings. The advantage of the CLI tool is that it can be invoked with different input values using a script and be left to complete in the background (say over a weekend). This could be particularly helpful for practitioners, as they can let the data be collected over time, while they tend to other tasks.
\subsection{Auxiliary Tools}\label{tool:auxillary}
The tool's package has two auxiliary CLI tools called the Verifier tool and the Recovery Tool at the user's disposal.\\
The \textbf{Verifier} tool can be used for validating results produced by the main tool. For instance, in cases where the dataset is too big to conduct multiple Classification tasks practically, one may choose to keep the Horizontal Expense to a fairly low value. This, however, would reduce the confidence in the results, since they may have overfitted a small portion of the large dataset. The Verifier tool can be invoked to try a (sub)set of Attribute Sets reported to be providing encouraging results, over the complete dataset, and observe the difference in the produced metrices between the two cases.\\

The \textbf{Recovery} tool is an SoS application that can be invoked to ``resume'' an abruptly stopped experiment due to mistakes (say by pressing ``Ctrl + C'') or problems (say power or network failures). When an experiment running for several hours or days gets killed, the Recovery tool can either dump the results already collected, or resume the experiment from the last saved state.
\section{Experimental Results}
In order to evaluate our model, and test the tool over a real world dataset, we conducted three sets of experiments over varying dataset sizes. For some experiments, we used two classification techniques as the learning mechanism - Naive Bayes and J48 Decision Trees. For larger experiments, we choose only J48 due to computational constraints. We now present the results of our experiments. The first two datasets are standard datasets popularly used for reporting results by researchers in the privacy domain. The third dataset is a real world credit card fraud dataset, with additional challenges. The tool was set throughout to use 5-fold cross-validation for producing any result.
\begin{figure*}[ht]
\centering 
    \subfigure[J48 Classifier]{
        \label{fig:results-adult-complete-j48}
        \includegraphics[width=0.47\textwidth]{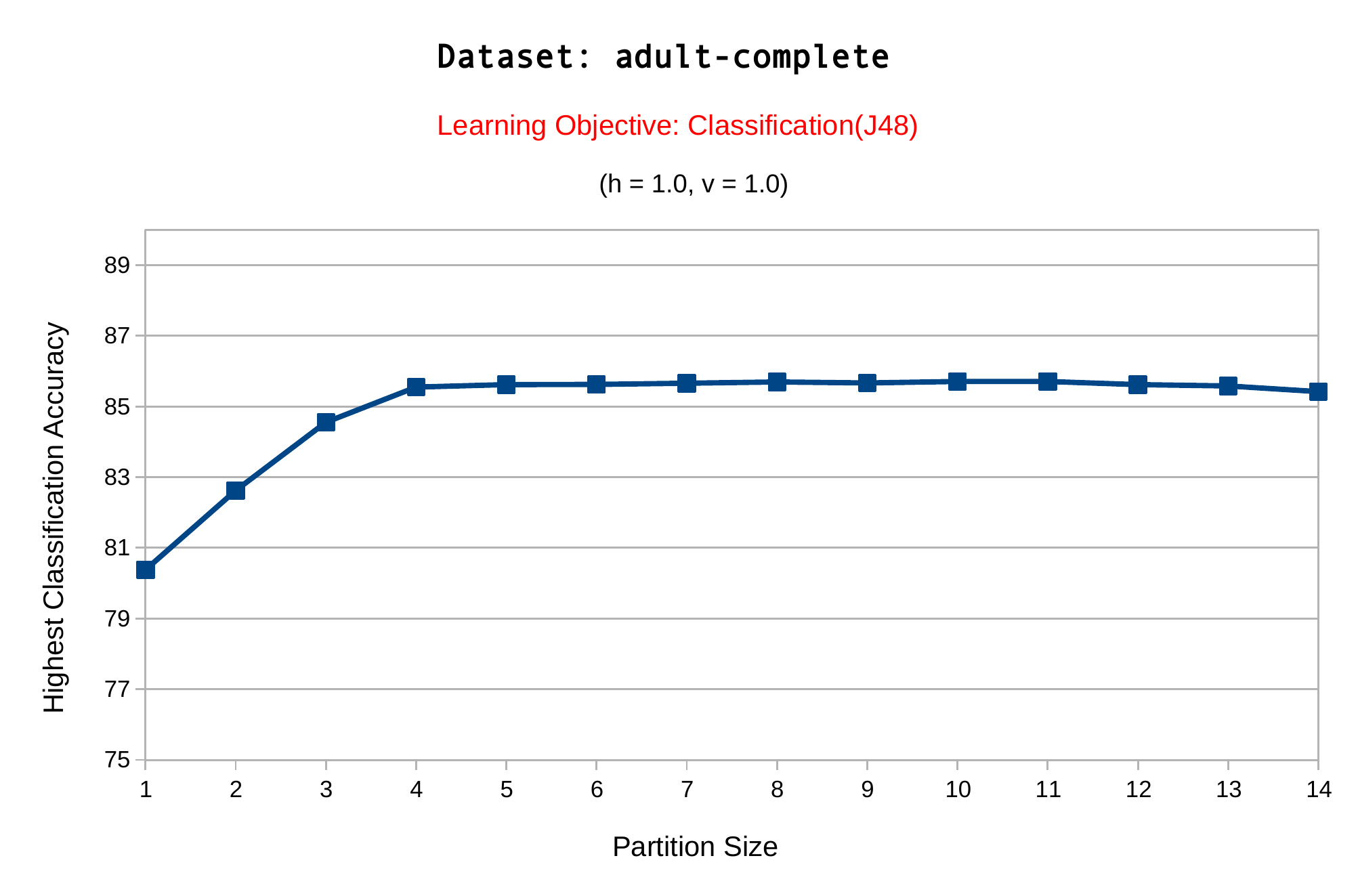}
    }
    \subfigure[Naive Bayes Classifier]{
        \label{fig:results-adult-complete-naivebayes}
        \includegraphics[width=0.47\textwidth]{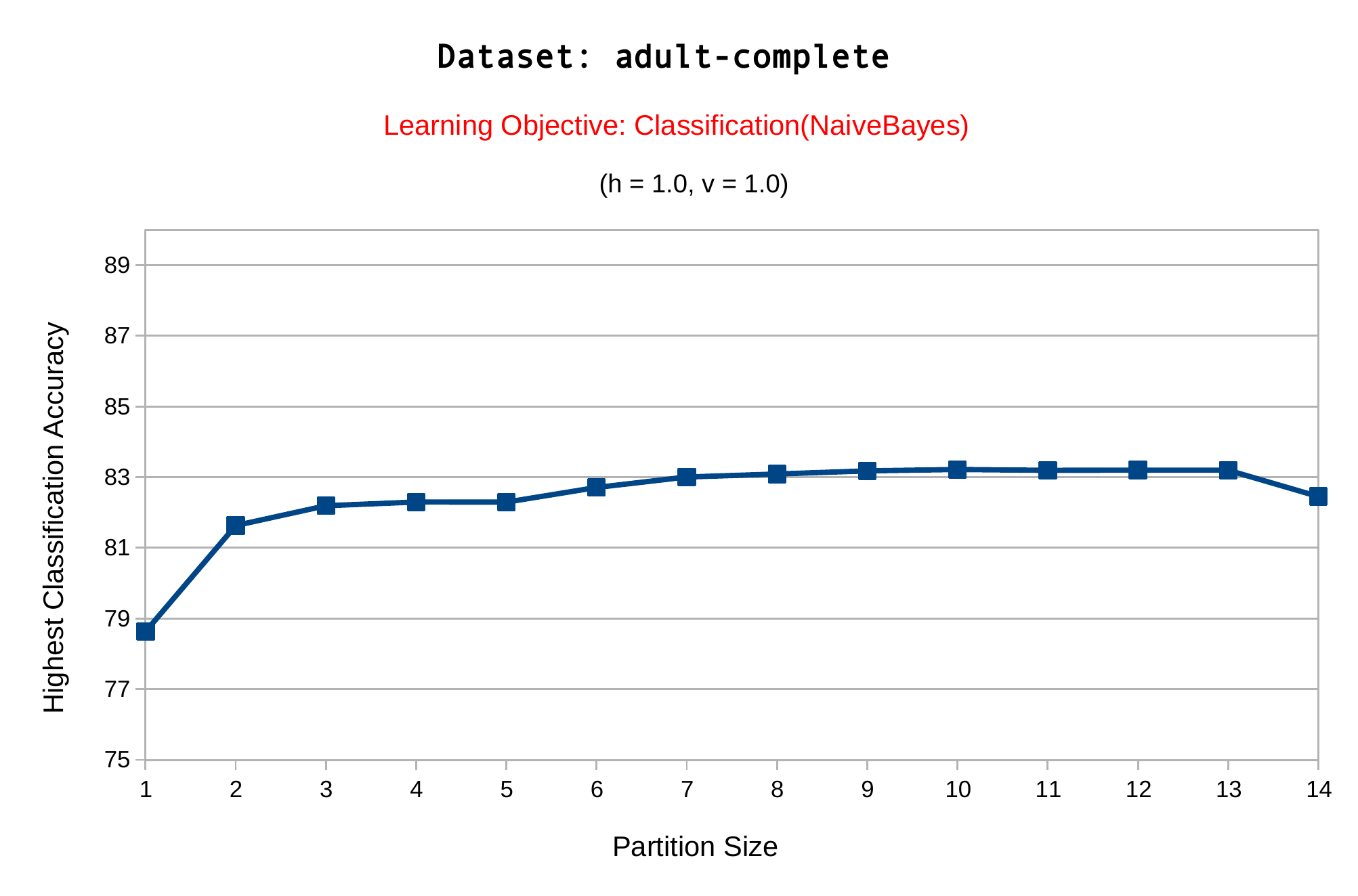}
    }
\caption{Highest Classification Accuracy achieved for \texttt{adult-complete} for varying values of partition size}
\label{fig:results-adult-complete}
\end{figure*}
\subsection{Truncated Adult dataset}\label{experiments-adult-iyengar}
The first objective of our experiments was to test if the hypothesis can even be applied to a real world dataset. We needed a small dataset to start with in order to quickly verify our model. To do this we chose a subset of UCI Adult Dataset, which was originally used by Iyengar\cite{iyengar2002transforming}. Over the years, the setup was replicated by many researchers (\cite{bayardo2005data}\cite{lefevre2006mondrian}\cite{lefevre2005incognito} etc.), and has become a de facto dataset to report results for researchers working in the privacy domain. We refer to this dataset as \texttt{adult-iyengar}. The dataset contained only 8 attributes (and the \textbf{class} attribute) from the original dataset. We skip the exact details of how the dataset was produced from the original dataset as they are exactly the same as described by Iyengar.

Figure \ref{fig:results-adult-iyengar} shows the results of the experiments. We intended to see how the classification accuracy was affected, on selecting a smaller subset of the dataset. For this, we varied the partition size from 1 to \textit{n} (in this case, \textit{n} = 8) and invoked the tool in order to find a partition, that provides accuracy within reasonable range of the reference accuracy value (accuracy achieved when the entire dataset is presented). It can be seen that the accuracy remains within reasonable limits for partition sizes of 3 or more. This means that choosing a smaller subset of attributes does not affect utility to a great extent, while providing a higher level of privacy.
\begin{figure*}[ht]
\centering 
    \subfigure[J48 Classifier with varying horizontal expense]{
        \label{fig:results-adult-complete-j48-variable-h}
        \includegraphics[height=0.19\textheight, width=0.95\textwidth]{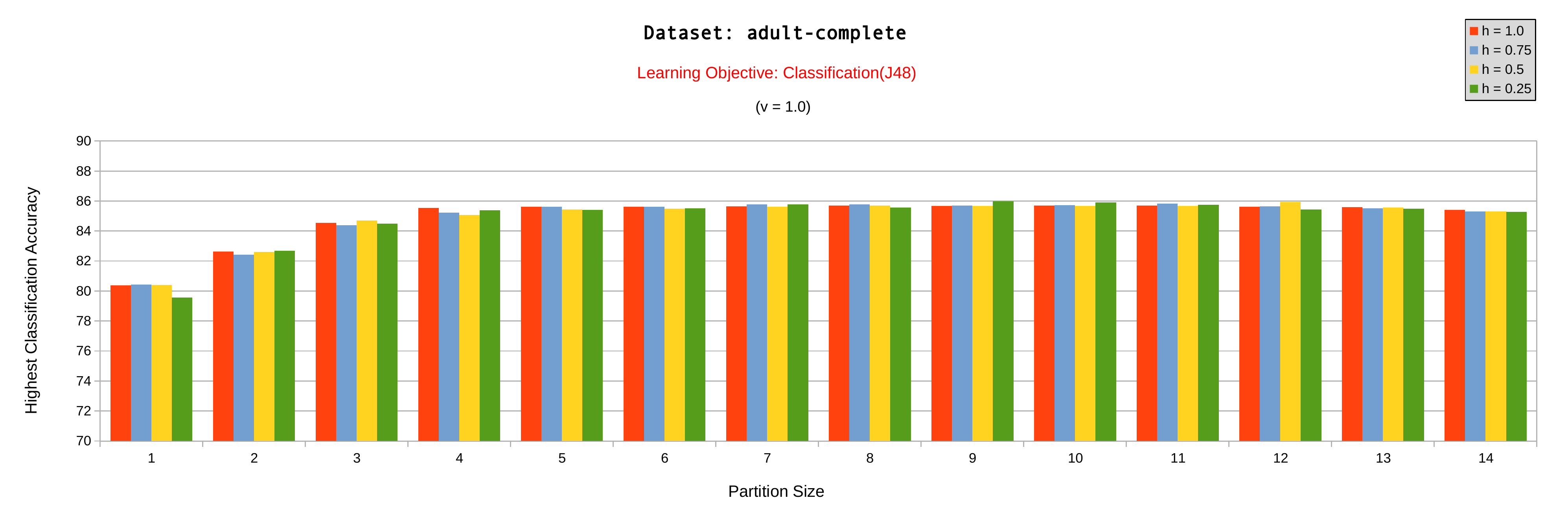}
    }
    \subfigure[J48 Classifier with varying vertical expense]{
        \label{fig:results-adult-complete-j48-variable-v}
        \includegraphics[height=0.19\textheight, width=0.95\textwidth]{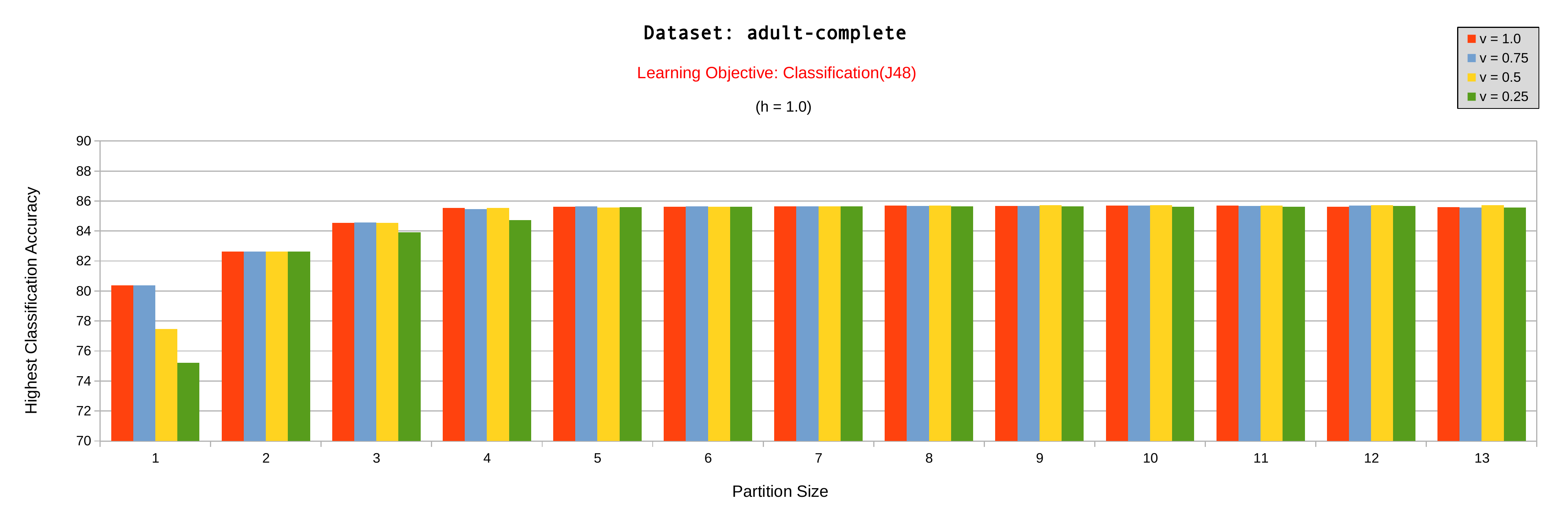}
    }
    \subfigure[Naive Bayes Classifier with varying horizontal expense]{
        \label{fig:results-adult-complete-naivebayes-variable-h}
        \includegraphics[height=0.19\textheight, width=0.95\textwidth]{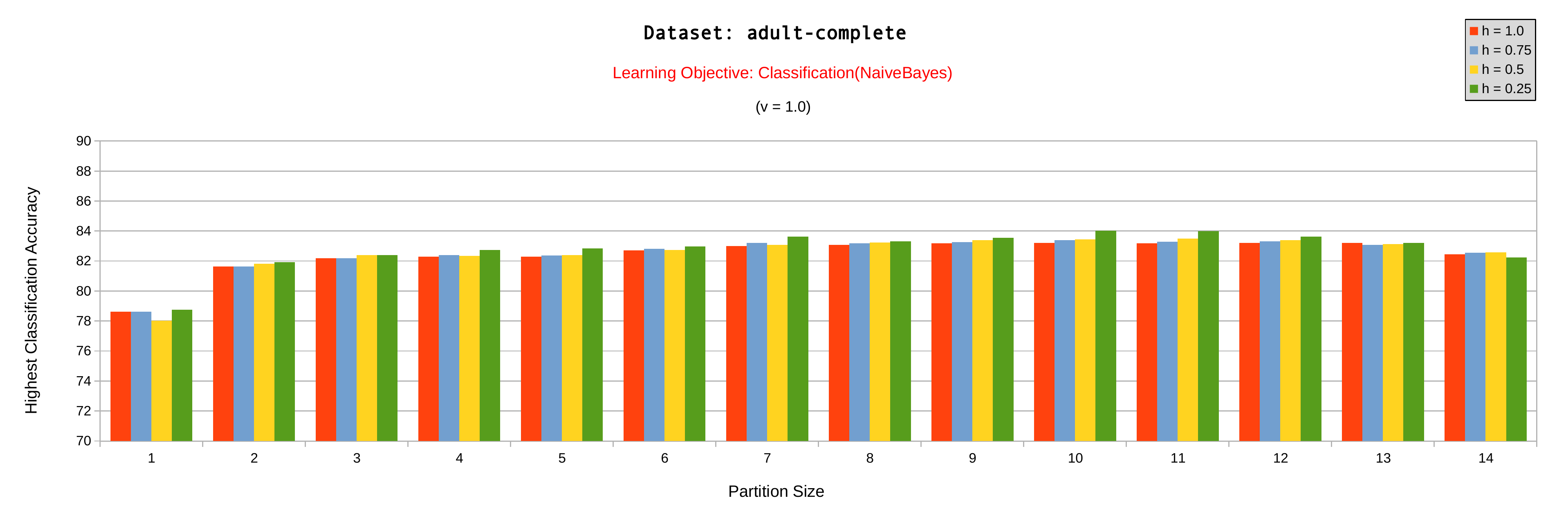}
    }
    \subfigure[Naive Bayes Classifier with varying vertical expense]{
        \label{fig:results-adult-complete-naivebayes-variable-v}
        \includegraphics[height=0.19\textheight, width=0.95\textwidth]{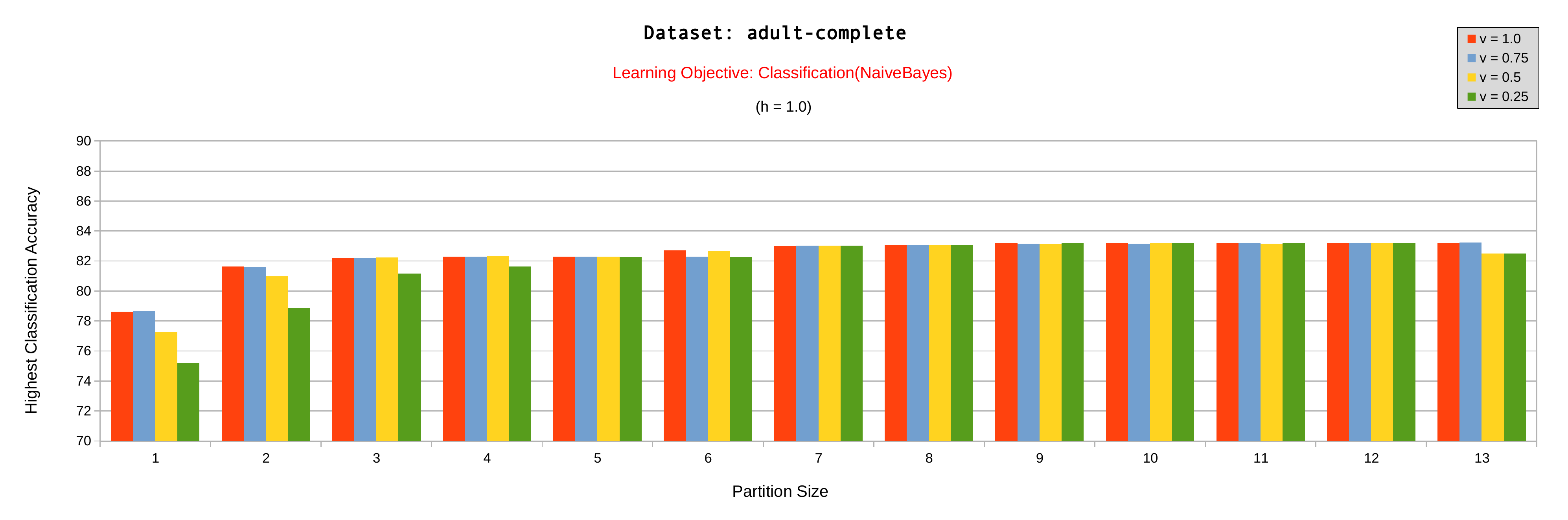}
    }
\caption{Highest Classification Accuracy achieved for \texttt{adult-complete} for varying values of partition size, vertical expense and horizontal expense}
\label{fig:results-adult-complete-variable-v-and-h}
\end{figure*}
\subsection{Complete Adult dataset}\label{experiments-adult-complete}
The \texttt{adult-iyengar} dataset can be considered a small, representative dataset, but it is certainly not exhaustive. Our next set of experiments used the complete UCI Adult dataset\cite{Lichman:2013}. First, we ran the same profile of experiments as we did before, i.e. we varied the values of partition size, and plotted the highest accuracy achieved for each partition size. During this phase, we kept the vertical and horizontal expense parameters at their full values, and asked the tool to use dictionary order generation instead of random generation for these experiments\footnote{The same profile applied to the experiments in Section \ref{experiments-adult-iyengar} as well.}. We refer to this dataset as \texttt{adult-complete}. 
The dataset contained all the 14 attributes (and the \textbf{class} attribute) of the original dataset. We removed any duplicate or conflicting instances, as well as those with missing values to obtain a clean dataset with 45,175 rows. The results of this set of experiments are shown in Figure \ref{fig:results-adult-complete}.
\begin{table*}[t]
    \centering
     \begin{tabular}{l p{30mm} l p{30mm}} 
     \hline
		 \multicolumn{4}{c}{\textbf{Attribute Set - \texttt{U}}} \\
	 \hline
		\textbf{Accuracy} & 99.88017 & \textbf{Time taken} & 1.09846 \\
		\textbf{TP\_1} & 0.52000 & \textbf{TP\_0} & 0.99964 \\
		\textbf{FP\_1} & 0.00035 & \textbf{FP\_0} & 0.48000 \\
		\textbf{FN\_1} & 0.48000 & \textbf{FN\_0} & 0.00035 \\
		\textbf{Precision\_1} & 0.72222 & \textbf{Precision\_0} & 0.99915 \\
		\textbf{Recall\_1} & 0.52000 & \textbf{Recall\_0} & 0.99964 \\
		\textbf{aROC\_1} & 0.77595 & \textbf{aROC\_0} & 0.77595 \\
		\textbf{aPR\_1} & 0.46787 & \textbf{aPR\_0} & 0.99915 \\
	\hline
     \end{tabular}
    \caption{Results for the \texttt{credit-card} for \texttt{(put number = 1)}}
    \label{table:results-creditcard-put-1}
\end{table*}
\begin{table*}[h]
  \centering
  \begin{tabular}{l p{23mm} p{23mm} p{23mm} p{23mm} p{23mm}}
    \hline
    & \multicolumn{5}{c}{\textbf{Attribute Set}} \\
    \hline
                          & \texttt{\{1, 2, 5, 8, 10, 11, 14, 15, 16, 17, 21, 25, 26, 28, 29\}}  & \texttt{\{1, 4, 5, 7, 10, 11, 13, 14, 15, 16, 17, 21, 25, 29, 30\}}  & \texttt{\{1, 2, 4, 5, 6, 7, 14, 17, 18, 19, 22, 27, 28, 29, 30\}}  & \texttt{\{3, 4, 5, 6, 7, 8, 10, 14, 16, 17, 19, 26, 27, 28, 30\}}  & \texttt{\{2, 5, 6, 10, 13, 14, 15, 16, 17, 19, 21, 22, 25, 26, 30\}}  \\
    \textbf{Accuracy}     & 99.95841
       & 99.95136
       & 99.95136
       & 99.95523
       & 99.95206
       \\
    \textbf{aROC\_1}      & 0.89360
         & 0.87870
               & 0.87207
               & 0.88925
               & 0.88382
               \\
    \textbf{aPR\_1}      & 0.76068
         & 0.75866
               & 0.75325
               & 0.75057
               & 0.75031
               \\
    \hline
    \\
    \end{tabular}
  \caption{Results for \texttt{credit-card} for \texttt{(put number = 0)}} \label{table:results-creditcard-put-0} 
\end{table*}
As seen, the trend observed for the experiments in Section \ref{experiments-adult-iyengar} is reproduced on a larger dataset. Beyond a certain partition size, the accuracy achieved is within a fair range of that being achieved with the use of all attributes. This provides more confidence that such a model can be used in a real world scenario.

We then performed two other sets of experiments as well on the \texttt{adult-complete} dataset. The experiments aimed at seeing the impact of considering only a smaller set of attributes and/or instances on the results of the tool. This is important because if the results degrade significantly on application of our engineering parameters, i.e. vertical and horizontal expense, then the tool cannot be considered practical enough to be used in a real world setting. The results are shown in Figure \ref{fig:results-adult-complete-variable-v-and-h}.

We changed the values of \textbf{v} and \textbf{h} to see if the trend is changed. Figure \ref{fig:results-adult-complete-j48-variable-h} and \ref{fig:results-adult-complete-naivebayes-variable-h} show the impact of varying the values of \textbf{h}, essentially meaning inclusion of lesser number of rows per partition. It can be seen that for both the classifiers, varying value of \textbf{h}, still keeps the accuracy within reasonable limits of the corresponding case, where all the rows were included in the partition. This shows that reducing the size of the partitions did not have significant impact on the model. This is reassuring because for large datasets, keeping a relatively low value of \textbf{h} could be a necessity, but doing so should not impact the overall results by much.

Lastly, we compared the accuracy readings reported by the tool for different values of \textbf{v}, essentially meaning that certain partitions were never tried by the tool, before reporting the highest accuracy for that partition number. Figure \ref{fig:results-adult-complete-j48-variable-v} and \ref{fig:results-adult-complete-naivebayes-variable-v} show the results of the experiments. One thing to point out here is that reducing \textbf{v} for partition size equal to \textit{n} (14 in this case) is not applicable, since there is only one partition to try in such cases. Clearly, the reduction of vertical expense parameter did have some impact on the highest accuracy, but the tool was still able to find a partition that could provide accuracy within a reasonable range of the corresponding experiment where all partitions were tried. 

This provides confidence in cases where the total number of partitions that can be formulated are too many. Trying out the learning objective even over a small set of systematically chosen partitions, can still provide a solution within practical limits for use.
\subsection{Credit Card Fraud dataset}\label{experiments-creditcard}
In order to evaluate the model and the tool we built over a real world dataset, we chose a credit card fraud dataset\cite{ccfdataset}. The dataset contains anatomized credit card transactions by some European cardholders in September 2013. We refer to this dataset as the \texttt{credit-card} dataset. The dataset has already been sanitized into 28 attributes (numbered \texttt{V1} to \texttt{V28}) along with two more attributes, \texttt{Amount} and \texttt{Time}, making a total of 30 attributes (plus the class attribute which has binary values with \texttt{1} indicating a fraudulent transaction and \texttt{0} indicating otherwise). The dataset is highly unbalanced\cite{dal2015calibrating}, which is typical for a credit card fraud use-case. Out of a total of 284,807 transactions, only 492 are tagged as frauds. This places an additional challenge since classification accuracy may no longer be a good metric for the model to use. We therefore, looked at other metrics that the tool generates in order to analyse the privacy concerns.\\
We now elaborate some of the used conventions:
\begin{itemize}[leftmargin=*]
    \item The tool numbered the attributes serially, in the order they are present in the input (ARFF\cite{arff}) data file.
    \item The tool produced a CSV file as output, containing the following fields for every classification task taken up in the experiment. The values of \textit{i} were \texttt{0} (the class of valid transactions) and \texttt{1} (the class of fraudulent transactions).

     \begin{tabular}{l p{53mm}} 
     \textbf{Attribute set} & The set of attributes tried for the task \\
     \textbf{Time taken} & The time taken, in seconds, for the task to complete \\
     \textbf{Accuracy} & The classification accuracy, in percentage, achieved for the task \\
     \textbf{TP\_\textit{i}} & The True Positives Rate, on scale of 0 to 1, for class \textit{i} \\
     \textbf{FP\_\textit{i}} & The False Positives Rate, on scale of 0 to 1, for class \textit{i} \\
     \textbf{FN\_\textit{i}} & The False Negatives Rate, on scale of 0 to 1, for class \textit{i} \\
     \textbf{Precision\_\textit{i}} & The Precision, on scale of 0 to 1, for class \textit{i} \\
     \textbf{Recall\_\textit{i}} & The Recall, on scale of 0 to 1, for class \textit{i} \\
     \textbf{aROC\_\textit{i}} & The Area under the ROC Curve, on scale of 0 to 1, for class \textit{i} \\
     \textbf{aPR\_\textit{i}} & The Area under the PR Curve, on scale of 0 to 1, for class \textit{i} \\
     \end{tabular}

    \item The \texttt{Time} and \texttt{Amount} attributes are the first and last attributes in the dataset, being referred to as \texttt{Attribute\# 1} and \texttt{Attribute\# 30} respectively. The attributes \texttt{V1} to \texttt{V28} are referred to as \texttt{Attribute\# 2} to \texttt{Attribute\# 29} respectively.
    \item Since we are more interested in the ability of the classification task to be able to predict a fraudulent transaction, throughout, we focus on the metrics collected for class \texttt{1} more than that for class \texttt{0}.
    \item Although the tool reports all the above metrics in the output file, for the discussion, we focus on the \textbf{area under PR curve}, \textbf{area under ROC curve} and the \textbf{Accuracy} as the main metrics for analysis. In particular, for credit card fraud datasets, which are often highly imbalanced in nature, using \textbf{area under PR curve} could be a better option than \textbf{area under ROC curve} or \textbf{Accuracy}\cite{creditcardmetrics}. Therefore, for discussions, we sort our results in non-decreasing order, with \textbf{aPR\_1}, \textbf{aROC\_1} and \textbf{Accuracy} being the primary, secondary and tertiary criterion respectively.
\end{itemize}
Based on the trade-off model, we treat two cases as our reference cases. The case where all attributes in the dataset are used, is the case where we provide least weightage to privacy, and highest weightage to utility. The exact opposite, is the case where we use only one attribute from the dataset at a time and try to perform the learning (in this case, perform a ``classification'' task using the ``J48'' decision tree algorithm). We present the findings of these two cases first to give an idea of the two extremes.
\begin{table*}[ht]
  \centering
  \begin{tabular}{l p{23mm} p{23mm} p{23mm} p{23mm} p{23mm}}
    \hline
    & \multicolumn{5}{c}{\textbf{Attribute Set}} \\
    \hline
                          & \texttt{\{1\}}  & \texttt{\{2\}}  & \texttt{\{3\}}  & \texttt{\{4\}}  & \texttt{\{5\}}  \\
    \textbf{Accuracy}     & 99.83083       & 99.82378       & 99.90131       & 99.85197       & 99.79558       \\
    \textbf{aROC\_1}      & 0.48328         & 0.49995               & 0.47138               & 0.48092               & 0.48616               \\
    \textbf{aPR\_1}      & 0.00162         & 0.00176               & 0.00092               & 0.00142               & 0.00197               \\
    \hline
    & \multicolumn{5}{c}{\textbf{Attribute Set}} \\
    \hline
                          & \texttt{\{6\}}  & \texttt{\{7\}}  & \texttt{\{8\}}  & \texttt{\{9\}}  & \texttt{\{10\}} \\
    \textbf{Accuracy}     & 99.82378       & 99.86607       & 99.83083       & 99.85902       & 99.85197       \\
    \textbf{aROC\_1}      & 0.49995         & 0.47890               & 0.48328               & 0.65941               & 0.79118               \\
    \textbf{aPR\_1}      & 0.00176         & 0.00127               & 0.00162               & 0.06737               & 0.32891               \\
    \hline
    & \multicolumn{5}{c}{\textbf{Attribute Set}} \\
    \hline
                          & \texttt{\{11\}} & \texttt{\{12\}} & \texttt{\{13\}} & \texttt{\{14\}} & \texttt{\{15\}} \\
    \textbf{Accuracy}     & 99.90836        & 99.92246        & 99.85197        & 99.92246        & 99.88017        \\
    \textbf{aROC\_1}      & 0.75380         & 0.73174         & 0.48092         & 0.92222         & 0.46470         \\
    \textbf{aPR\_1}      & 0.48456         & 0.41327         & 0.00142         & 0.57707         & 0.00112         \\
    \hline
    & \multicolumn{5}{c}{\textbf{Attribute Set}} \\
    \hline
                          & \texttt{\{16\}}  & \texttt{\{17\}}  & \texttt{\{18\}}  & \texttt{\{19\}}  & \texttt{\{20\}}  \\
    \textbf{Accuracy}     & 99.86607        & 99.85902        & 99.90131        & 99.83787        & 99.83787        \\
    \textbf{aROC\_1}      & 0.75253         & 0.80987         & 0.64286         & 0.47388         & 0.47388         \\
    \textbf{aPR\_1}      & 0.52713         & 0.44304         & 0.36259         & 0.00153         & 0.00153         \\
    \hline
    & \multicolumn{5}{c}{\textbf{Attribute Set}} \\
    \hline
                          & \texttt{\{21\}}  & \texttt{\{22\}}  & \texttt{\{23\}}  & \texttt{\{24\}}  & \texttt{\{25\}} \\
    \textbf{Accuracy}     & 99.86607        & 99.83787        & 99.86607        & 99.83787        & 99.83083        \\
    \textbf{aROC\_1}      & 0.47890         & 0.47388         & 0.47890         & 0.47388         & 0.48328         \\
    \textbf{aPR\_1}      & 0.00127         & 0.00153         & 0.00127         & 0.00153         & 0.00162         \\
    \hline
    & \multicolumn{5}{c}{\textbf{Attribute Set}} \\
    \hline
                          & \texttt{\{26\}} & \texttt{\{27\}} & \texttt{\{28\}} & \texttt{\{29\}} & \texttt{\{30\}} \\
    \textbf{Accuracy}     & 99.78853        & 99.85197        & 99.78149        & 99.78149        & 99.84492        \\
    \textbf{aROC\_1}      & 0.49995         & 0.48092         & 0.48706         & 0.48706         & 0.47272         \\
    \textbf{aPR\_1}      & 0.00211         & 0.00142         & 0.00213         & 0.00213         & 0.00147         \\
  \end{tabular}
  \caption{Results for \texttt{credit-card} for \texttt{(put number = -1)}} \label{table:results-creditcard-put--1} 
\end{table*}

The data produced by the tool when invoked with \texttt{put number = -1} (\texttt{partition size = 1}) is shown in Table \ref{table:results-creditcard-put--1}. The tool produced values for all fields shown in Table \ref{table:results-creditcard-put-1}, but as mentioned before, we have shown only a fraction of overall data, that is relevant. Clearly, the \textbf{Accuracy} is very high in all cases. This can be attributed to the fact that since the fraudulent transactions are fairly low in number, even a dumb predictor which predicts the same class all the time (class \textit{0}), can still end up with very high Accuracy.
\begin{table*}[h]
  \centering
  \begin{tabular}{l p{23mm} p{23mm} p{23mm} p{23mm} p{23mm}}
     \multicolumn{6}{c}{\texttt{(put number = 0.75)}}
    \\
    \hline
    & \multicolumn{5}{c}{\textbf{Attribute Set}} \\
    \hline
                          & \texttt{U - \newline\{1, 3, 9, 13\}}  & \texttt{U - \newline\{1, 2, 27, 28\}}  & \texttt{U - \newline\{1, 4, 20, 23\}}  & \texttt{U - \newline\{1, 2, 11, 27\}}  & \texttt{U - \newline\{1, 2, 11, 28\}}  \\
    \textbf{Accuracy}     & 99.94607
       & 99.94924
       & 99.94431
       & 99.94536
       & 99.94572
       \\
    \textbf{aROC\_1}      & 0.87628
         & 0.86957
               & 0.85941
               & 0.86722
               & 0.87862
               \\
    \textbf{aPR\_1}      & 0.72319
         & 0.72218
               & 0.71789
               & 0.71620
               & 0.70504
               \\
    \hline
    \\
    \end{tabular}
    \begin{tabular}{l p{23mm} p{23mm} p{23mm} p{23mm} p{23mm}}
     \multicolumn{6}{c}{\texttt{(put number = 0.5)}}
    \\
    \hline
    & \multicolumn{5}{c}{\textbf{Attribute Set}} \\
    \hline
                          & \texttt{U - \newline\{1, 4, 7, 12, 13, 15, 26, 30\}}  & \texttt{U - \newline\{2, 7, 8, 9, 12, 19, 26, 27\}}  & \texttt{U - \newline\{2, 6, 9, 12, 15, 25, 27, 28\}}  & \texttt{U - \newline\{1, 2, 3, 5, 12, 19, 28, 30\}}  & \texttt{U - \newline\{5, 13, 17, 23, 26, 27, 28, 30\}}  \\
    \textbf{Accuracy}     & 99.95030       & 99.95171       & 99.95277       & 99.95418       & 99.95136       \\
    \textbf{aROC\_1}      & 0.89460         & 0.87972               & 0.88960               & 0.88286               & 0.90483              \\
    \textbf{aPR\_1}      & 0.76495         & 0.75878               & 0.75770               & 0.75513               & 0.75484               \\
    \hline
    \\
    \end{tabular}
    \begin{tabular}{l p{23mm} p{23mm} p{23mm} p{23mm} p{23mm}}
     \multicolumn{6}{c}{\texttt{(put number = 0.25)}}
    \\
    \hline
    & \multicolumn{5}{c}{\textbf{Attribute Set}} \\
    \hline
                          & \texttt{U - \newline\{1, 3, 6, 7, 8, 11, 12, 15, 18, 19, 20\}}  
                          & \texttt{U - \newline\{1, 4, 6, 9, 15, 20, 21, 22, 23, 25, 28\}}  
                          & \texttt{U - \newline\{4, 5, 6, 9, 11, 12, 18, 20, 23, 28, 29\}}  
                          & \texttt{U - \newline\{1, 4, 5, 7, 10, 13, 16, 25, 26, 27, 29\}}  
                          & \texttt{U - \newline\{1, 2, 6, 9, 14, 20, 22, 23, 24, 27, 29\}}  \\
    \textbf{Accuracy}     & 99.95241       & 99.94536       & 99.95136       & 99.94184       & 99.94008       \\
    \textbf{aROC\_1}      & 0.87247         & 0.86184               & 0.86064               & 0.85308               & 0.85677               \\
    \textbf{aPR\_1}      & 0.73969         & 0.71005               & 0.70538               & 0.69688               & 0.69353               \\
    \hline
    \end{tabular}
  \caption{Results for \texttt{credit-card} for \texttt{(0 < put number < 1)}} \label{table:results-creditcard-put-gt0} 
\end{table*}
For all other cases, where the number of partitions tried were fairly high in number, we followed the following procedure, to narrow down to a small set of useful partitions, for a given put number:
\begin{enumerate}[leftmargin=*]
    \item We first ran a longer, initial experiment, for the given put number, with small values of vertical and horizontal expenses (\textbf{v} and \textbf{h}) to avoid high computational costs.
    \item We then sorted the collected results using sort criteria mentioned before. In order to validate these results over the complete dataset, we chose a small set (in the order of 20 to 100) of partitions which performed well in the initial experiment, and ran our \textbf{Verifier} CLI tool (\ref{tool:auxillary}) over these set of partitions, to collect statistics over the complete dataset. This alleviated our concerns of overfitting.
    \item We sorted the collected results again using the sort criteria mentioned before. We then chose the top five partitions that performed the best for the given \texttt{put number}.
\end{enumerate}
The case which falls in the middle of our privacy model, is where \texttt{put number} is chosen as \texttt{0}. In this case, it mapped to a partition size of \texttt{15} (out of \texttt{30}). The top five results, obtained via the selection procedure mentioned before are shown in Table \ref{table:results-creditcard-put-0}. We must add that we were able to come up with these sets of partitions, despite the fact that we had to use fairly low values of \textbf{v} and \textbf{h}, in order to keep the number of partitions to be tried, to a manageable number. This provides a good confidence for using the tool, for cases where computational resources are limited.

The cases where utility were given higher precedence over privacy were the ones where the \texttt{put number} was set to a value ``relatively'' closer to \texttt{1} as compared to \texttt{-1}. For our experiments, we chose 3 values for usage: \texttt{0.75}, \texttt{0.5} and \texttt{0.25}. They mapped to \texttt{partition size} of \texttt{26}, \texttt{22} and \texttt{19} respectively. Selected results for these values are shown in Table \ref{table:results-creditcard-put-gt0}. As discussed before, due to lack of space, we are presenting only selected metrics, that too for the top five partitions only. For theses cases, we use a compact representation of a partition, where we show the indices of the attributes that should be removed from the original set containing all the attributes (\texttt{U}), instead of listing the attributes forming the partition.

It can be seen that the partitions for these cases actually perform even better than the case when we used the whole dataset. This behaviour is typical for many feature selection activities, where dropping least relevant features (which are essentially noise for the learning process) can actually increase the efficiency of the learning component.
\begin{table*}[ht]
  \centering
  \begin{tabular}{l p{23mm} p{23mm} p{23mm} p{23mm} p{23mm}}
     \multicolumn{6}{c}{\texttt{(put number = -0.25)}}
    \\
    \hline
    & \multicolumn{5}{c}{\textbf{Attribute Set}} \\
    \hline
                          & \texttt{\{3, 7, 9, 10, 13, 14, 16, 17, 20, 21, 23\}}  & \texttt{\{3, 4, 5, 10, 14, 15, 16, 17, 18, 23, 30\}}  & \texttt{\{3, 4, 10, 14, 16, 17, 19, 23, 27, 29, 30\}}  & \texttt{\{1, 4, 10, 14, 15, 16, 17, 19, 21, 24, 27\}}  & \texttt{\{4, 5, 7, 9, 10, 14, 17, 21, 23, 25, 30\}}  \\
    \textbf{Accuracy}     & 99.95206
       & 99.94889
       & 99.95453
       & 99.95136
       & 99.94713
       \\
    \textbf{aROC\_1}      & 0.89486
         & 0.88855
               & 0.86523
               & 0.90252
               & 0.88281
               \\
    \textbf{aPR\_1}      & 0.75701
         & 0.75214
               & 0.75201
               & 0.75118
               & 0.75087
               \\
    \hline
    \\
    \end{tabular}
    \begin{tabular}{l p{23mm} p{23mm} p{23mm} p{23mm} p{23mm}}
     \multicolumn{6}{c}{\texttt{(put number = -0.5)}}
    \\
    \hline
    & \multicolumn{5}{c}{\textbf{Attribute Set}} \\
    \hline
                          & \texttt{\{2, 5, 10, 11, 14, 15, 16, 30\}}  & \texttt{\{8, 10, 12, 14, 20, 22, 24, 30\}}  & \texttt{\{6, 7, 13, 14, 15, 20, 29, 30\}}  & \texttt{\{3, 8, 11, 16, 17, 21, 26, 30\}}  & \texttt{\{7, 11, 14, 16, 22, 24, 26, 29\}}  \\
    \textbf{Accuracy}     & 99.95065       & 99.94748       & 99.94290       & 99.93655       & 99.94360       \\
    \textbf{aROC\_1}      & 0.87812         & 0.87601               & 0.88840               & 0.87209               & 0.88554              \\
    \textbf{aPR\_1}      & 0.74206         & 0.72595               & 0.72089               & 0.70121               & 0.70012               \\
    \hline
    \\
    \end{tabular}
    \begin{tabular}{l p{23mm} p{23mm} p{23mm} p{23mm} p{23mm}}
     \multicolumn{6}{c}{\texttt{(put number = -0.75)}}
    \\
    \hline
    & \multicolumn{5}{c}{\textbf{Attribute Set}} \\
    \hline
                          & \texttt{\{17, 23, 24, 25\}} 
                          & \texttt{\{16, 17, 22, 28\}} 
                          & \texttt{\{16, 17, 20, 27\}} 
                          & \texttt{\{17, 19, 27, 29\}} 
                          & \texttt{\{17, 19, 20, 23\}}  \\
    \textbf{Accuracy}     & 99.92034       & 99.92492       & 99.91646       & 99.92351       & 99.92175       \\
    \textbf{aROC\_1}      & 0.84071         & 0.86010               & 0.83149               & 0.82934              & 0.83944               \\
    \textbf{aPR\_1}      & 0.62026         & 0.61283               & 0.60366               & 0.60100               & 0.59993               \\
    \hline
    \end{tabular}
  \caption{Results for \texttt{credit-card} for \texttt{(-1 < put number < 0)}} \label{table:results-creditcard-put-lt0} 
\end{table*}

The cases where privacy were given more importance than utility, are the ones where we kept the \texttt{put number} below \texttt{0}. The values we chose were: \texttt{-0.25}, \texttt{-0.5} and \texttt{-0.75}. They mapped to \texttt{partition size} of \texttt{11}, \texttt{8} and \texttt{4} respectively. 

The top five results obtained via the selection procedure mentioned before are shown in Table \ref{table:results-creditcard-put-lt0}. It can be seen that even dropping to a value as low as \texttt{-0.75} still provides us with some partitions having decent values for the \textbf{aPR\_1} metric. For other two cases, the values for the \textbf{aPR\_1} metric is actually as good as cases with higher precedence to utility. This is a reminder to the fact that even a privacy model which is simply based on the \textit{Collection limitation} and \textit{Data minimization} practices\cite{cavoukian2009privacy}, can still be applicable to real world scenarios.

\section{Related Work}\label{related}
We provide a brief overview of some of the work, that is relevant to our own work in this section. We divide the work of our interest into three categories (not necessarily exclusive).

First, there has been significant amount of work which can be summed up as ``attempts to build differentially private\cite{Dwork:2006:DP:2097282.2097284} versions of certain machine learning techniques''. For instance, researchers have worked on building differentially private versions of SVM\cite{rubinstein2009learning}, linear and logistic regression\cite{zhang2012functional}, bayesian detection\cite{li2015privacy}, random forests\cite{fletcher2017differentially}, nearest neighbor classification\cite{gursoy2017differentially} and even the relatively recent deep learning methods\cite{abadi2016deep}. Attempts to articulate generic methods for doing so like ``input perturbation'' or ``output perturbation'' have also been made\cite{sarwate2013signal}. From a practitioner's view, these works can only be of importance if they are converted into (possibly free and open-source) software components, which can be integrated with either an existing application, or be used as a ``black-box''. Unfortunately, majority of them are restricted to demo applications only, and not as stable software components to be used in production. There are some appreciable initiatives towards bridging this gap \cite{mohan2012gupt}\cite{mcsherry2009privacy}\cite{roy2010airavat}\cite{katla2017dpweka}, but more of such efforts are required for software providers to easily integrate privacy solutions in the routine development cycle.

Second, substantial research has gone in achieving differential privacy\cite{Dwork:2006:DP:2097282.2097284} in a practical setting by using the general concept of \textit{anonymisation}. The most popular examples are the usage of techniques like ``generalisation'' and ``suppression'' to achieve the notion of \textit{k}-anonymity\cite{samarati1998protecting}. Although refinements of \textit{k}-anonymity, such as \textit{l}-diversity \cite{machanavajjhala2007diversity} and \textit{t}-closeness\cite{li2007t} have also been developed, it can be reasonable to say that \textit{k}-anonymity still remains a popular choice of researchers, with multiple versions of the same (datafly\cite{sweeney1998datafly}, mondrian\cite{lefevre2006mondrian}, incognito\cite{lefevre2005incognito} etc.) available for practical usage under the \textbf{UTD Anonymization Toolbox}\cite{utdtoolbox}. However, a practical limitation of \textit{k}-anonymity that we observed while trying to use the toolbox is that implementing generalisation involves a non trivial task of defining a ``taxonomy tree'' for nominal attributes, so that the algorithm can move up in the tree to replace specific values (say for an attribute called \textit{workclass}) like ``Local Government'', ``Federal Government'' or ``State Government'' to a more general value ``Government''. \footnote{This particular example is quoted from the work by Iyengar \cite{iyengar2002transforming}.} If two researchers select two different ways to generalise values, we may end up with two different, incomparable versions of the same dataset, both implementing the same version of \textit{k}-anonymity with same value of \textit{k}. From a practitioner's view, providing these taxonomies manually will involve the arduous process of agreement among the stakeholders.

Third, there are relatively recent works, which are precursor to our work, that propose the idea of \textit{privacy-aware} feature selection \cite{pattuk2015privacy}\cite{pattuk2016optimizing}\cite{aristodimou2016privacy}\cite{yang2014differentially}\cite{pmlr-v30-Guha13}\cite{sheikhalishahi2017privacy}. Our work contrasts these works in either one or both of the following ways. First, our model doesn't expect any prerequisites from the user in the fields of Probability, Statistics, Information Theory etc. We do so by explaining the nuances of the model using metrics that a practitioner may already be accustomed to (such as ``Classification Accuracy''), instead of defining a new privacy metric that needs background reading. Second, we chose to provide an engineering solution to the problem, by providing a hands-on tool to analyze our model over any dataset, with no or minimal pre-processing\footnote{Our tool provides one-click options to handle missing values or duplicate instances in the input dataset, providing a safety net for algorithms that can't work with them.}. This too is important, because we intend to provide a solution for a community which might be tempted towards using a GUI based tool for analysing their dataset for impacts of privacy preservation. To the best of our knowledge, our work is the first to address these issues.
\section{Conclusion and Future Work}
In this work, we presented the idea, prototype and initial results related to a tool intended to aid software practitioners in making crucial decisions about usage of user data in their applications. In order to do so, we picked a fairly simple privacy model, which can be appreciated without any background knowledge in Data Science.

We showed an evaluation of our model via the use of our tool with experiments designed over the UCI Adult Dataset, a de-facto benchmark for researchers in the field. We also showed how the tool can be used in a real world scenario, by running various experiments over a credit card fraud dataset containing actual transaction data of some European cardholders.

The usual way of using this tool should be starting with a fairly low value of \texttt{put number} (or \texttt{partition size}) and evaluate the collected metrics for the given choice. If the results are within ``acceptable'' range of utility (e.g. \textbf{Accuracy} in excess of \texttt{85\%} or \textbf{area under PR Curve} in excess of \texttt{0.75}), a suitable partition (or set of partitions) shall be tried out in real world, and be used for production. Otherwise, another experiment, with slightly higher value of \texttt{put number} should be tried, and so on.

In future versions, we expect to make it more generic, by including features such as support for Regression and Clustering techniques as well. We plan to add support for choosing from a wide range of learning objectives, as well as provide a mechanism to link the tool to \textit{any} learning component in general, via a standardized call-and-return interface.
\bibliographystyle{ACM-Reference-Format}
\bibliography{bibliography}

\end{document}